\documentstyle[epsfig]{aipproc}        
\newcommand{\gnufig}[3]{
\begin{figure}[t]
\begin{center}
{#3}
\end{center}
\caption{#2}
\label{#1}
\end{figure}}

\def\be{\begin{equation}}
\def\ee{\end{equation}}
\def\bea{\begin{eqnarray}}
\def\eea{\end{eqnarray}}

\begin{document}
\title{Is charm the key to understanding diffraction in DIS ? }
\author{M.F. McDermott}
\address{DESY, Notkestrasse~85,~D-22603~Hamburg,~Germany}
\maketitle
\begin{abstract}
This talk concerns the production of open charm in diffractive deep
inelastic scattering. This has been calculated recently in the context
of the semi-classical approach to diffraction. A comparison is made to
approaches in which the diffractive exchange is modelled by the
exchange of two gluons in the $t$-channel. Two phenomenological test of the
underlying partonic process are discussed.
\end{abstract}

The first results on the production of open charm in diffractive DIS
were reported by the H1 collaboration \cite{MFM:H1WAR}, for the 1994
running period, last summer. 
Results from H1 and Zeus were given at this conference
\cite{MFM:DATCHI}. The experimental signature is a particular decay
channel of $D^*$ mesons with a $p_{\perp}$ above $1$~GeV; so far about
30 events have been found in total but this is expected to increase to
at least 100 with the 1996 data. The heavy charm quark mass permits a reliable 
calculation within perturbative QCD. By examining exclusive
channels such as open charm production it is hoped that more can be
learned about the total diffractive sample in DIS.

The simplest QCD model one can think of for diffraction is the
exchange of two gluons in a colour singlet in the $t$-channel.
Several recent papers \cite{MFM:TWOG} present predictions for diffractive charm
which have this mechanism of diffractive exchange in  common. 
The leading order graph is shown in Fig.(\ref{F:MFM:TWOG}). 
The black blob in the figure represents the different ways in which higher order
corrections and the gluon propagators have been considered in 
the various approaches.
Unfortunately a complete ${\cal O}(\alpha_s)$ calculation, 
which would also include a gluon in the final state is not yet
available. In terms of the diffractive structure function, these 
corrections are expected to be important when the diffractive mass is
large compared to $Q^2$. More exclusive 
predictions for diffractive charm production  are only 
available for the pure $c \bar c$ diffractive
final state.  The differential cross sections for 
transversely and longitudinally 
polarized photons can be written, in the double leading logarithmic
approximation in terms of the square of the gluon density, $G(\xi)$,
as follows 
\bea
\frac{d^2 \sigma_L}{d \alpha dp_{\perp}^2} & = & \frac{2e_c^2\alpha_{em}\alpha_s^2
\pi^2[\xi G(\xi)]^2C}{3(a^2+p_\perp^2)^6}[\alpha(1-\alpha)]^2Q^2
(a^2-p_\perp^2)^2  \\
\frac{d^2 \sigma_T}{d\alpha dp_{\perp}^2} & = & \frac{e_c^2\alpha_{em}\alpha_s^2\pi^2
[\xi G(\xi)]^2C}{6(a^2+p_\perp^2)^6}\left[4(\alpha^2+(1-\alpha)^2)p_\perp^2
a^4+m_c^2(a^2-p_\perp^2)^2\right]  \\
a^2 & = & \alpha (1-\alpha) Q^2 + m_c^2 
\eea
\noindent where $\alpha$ is the momentum fraction of the virtual
photon carried by the quark, $p_{\perp}^2$ is its transverse momentum
squared  and  $\xi$  (or $x_{I\!\!P}$) is the
longitudinal momentum fraction lost by the elastically scattered proton.
The factor $C$ parameterizes the required extrapolation from 
$t\approx 0$ to the integrated cross section, $C \approx \Lambda_{QCD}^2$.

\gnufig{F:MFM:TWOG}{Two gluon exchange in the $t$-channel.}{
\vspace{0cm}
\hspace{1cm}
\epsfig{file=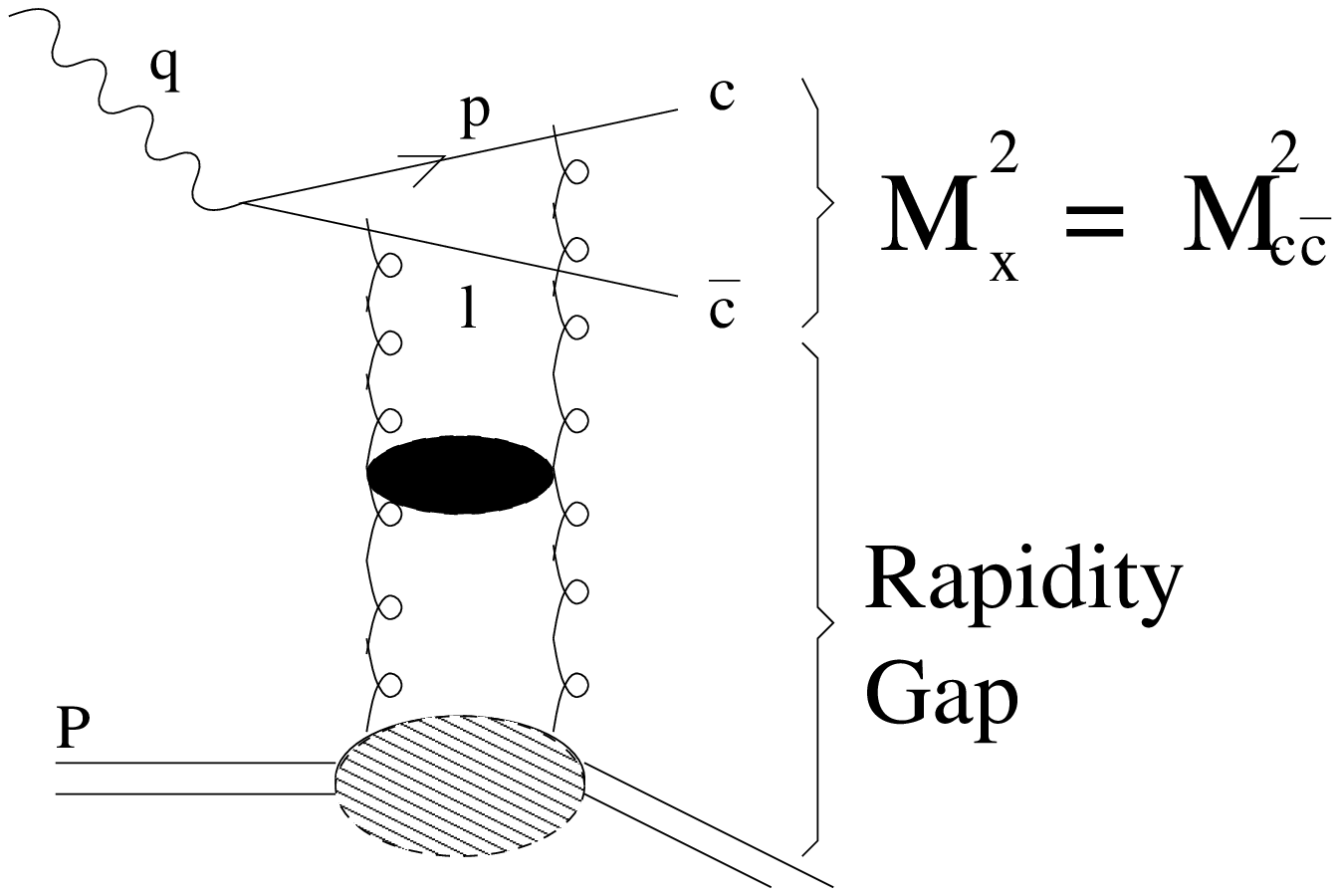,width=8cm,height=6cm}%
\hspace{2cm}
}

An alternative view of diffraction, known as the semiclassical
approach, which is very close in spirit to
the aligned jet model has been developed recently
\cite{MFM:BHM}. In this proton rest frame calculation, leading twist  
diffraction results from the scattering of asymmetric partonic
fluctuations of the virtual photon from the soft colour field of the
proton (i.e. those configurations containing a parton which carries only a
small fraction of the photon longitudinal momentum). The presence of
this `wee' or `slow' parton, which necessarily has a low $p_{\perp}$, allows
the fluctuation to develop a large transverse 
size by the time it arrives at the proton. 

\gnufig{F:MFM:QQG}{Dominant partonic process for diffractive
charm production in the semiclassical approach.}{
\hspace{0.7cm}
\epsfig{file=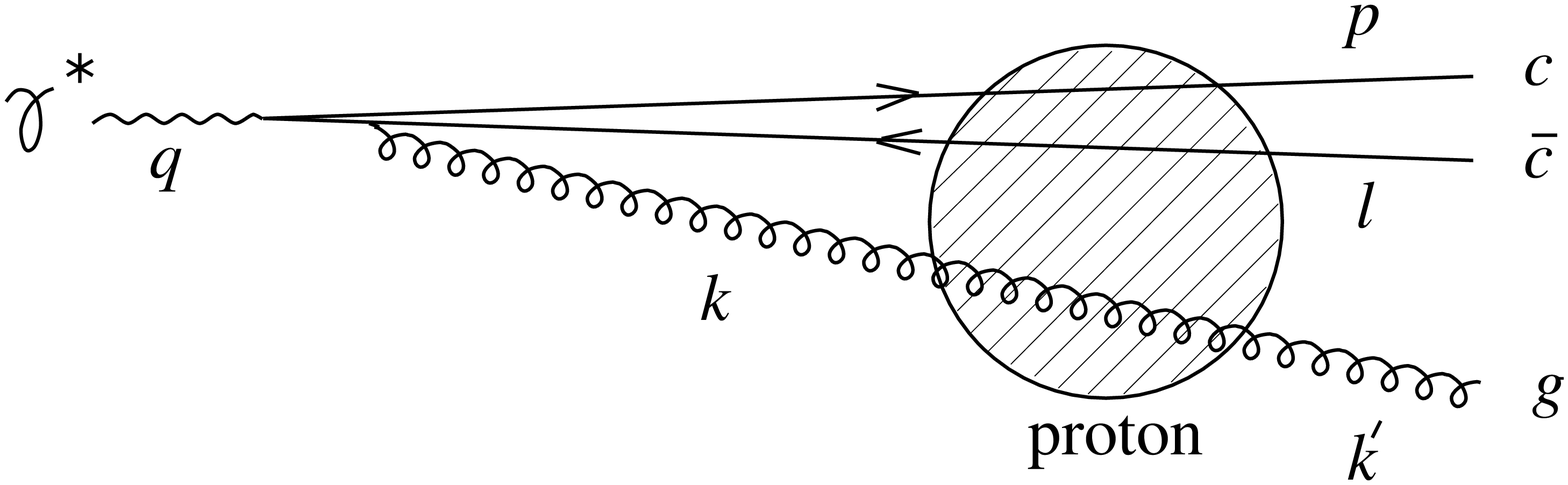,width=12cm}
}

Configurations in which the wee parton is a charm quark are suppressed
by powers of the charm mass, these include the contributions from the
leading order $c {\bar c}$ configuration and the corresponding QCD Compton graphs. 
The dominant contribution is ${\cal O}(\alpha_s)$ and contains a 
wee gluon (cf. Fig.(\ref{F:MFM:QQG})). 
Expressed in terms of the momentum fraction, $\alpha'$, and
transverse momentum, $k_{\perp}^{'2}$, of the gluon the differential cross sections
are given by \cite{MFM:BHMC}
\bea
\frac{d\sigma_L}{d\alpha dp_\perp^2d\alpha'dk_\perp'^2}&
=&\frac{e_c^2\alpha_{em}\alpha_s}{16\pi^2}\,\frac{\alpha'Q^2
p_\perp^2}{[\alpha(1\!-\!\alpha)]^2N^4}f^{\cal A}(\alpha'N^2,k_\perp')
\,, \nonumber\\
& & \\
\frac{d\sigma_T}{d\alpha dp_\perp^2d\alpha'dk_\perp'^2}& =&
\frac{e_c^2\alpha_{em}\alpha_s}{128\pi^2}\,\frac{\alpha'
\left\{[\alpha^2+(1\!-\!\alpha)^2]\,[p_\perp^4+a^4]+2p_\perp^2m_c^2\right\}}
{[\alpha(1\!-\!\alpha)]^4N^4} f^{\cal A}(\alpha'N^2,k_\perp')   \\
f^{\cal A}(\alpha'N^2,k_\perp') &=& \int_{x_\perp}\left|\int\frac{d^2k_\perp}
{(2\pi)^2}\left(\delta^{ij}+\frac{2k_\perp^ik_\perp^j}{\alpha'N^2}\right)
\frac{\mbox{tr}\tilde{W}^{\cal A}_{x_\perp}(k_\perp'\!-\!k_\perp)}
{\alpha'N^2+k_\perp^2}\right|^2 \,  \\
N^2&=&Q^2+\frac{p_\perp^2+m_c^2}{\alpha(1\!-\!\alpha)} \,.
\eea

The heavy charm mass ensures that the $c {\bar c}$-pair stays small 
in transverse space and behaves like a gluon terms of
colour. The superscript ${\cal A}$, which stands for adjoint,
is written here since we are effectively testing the protons 
field with two ``gluons''. The eikonal factor $W^{\cal A}$
parameterizes the interaction of the system of partons with the proton
and depends in general on the transverse momentum lost by the wee
parton. Taking its trace projects onto the colour singlet
configurations relevant for diffraction.
Integration over the final state variables
of the wee parton in the leading $\ln(1/x)$ approximation gives

\bea
\frac{d\sigma_L}{d\alpha dp_\perp^2} & = &\frac{e_c^2\alpha_{em}\alpha_s
\ln(1/x)h_{\cal A}}{2\pi^3(a^2+p_\perp^2)^4}\,[\alpha(1-\alpha)]^2Q^2 
p_\perp^2  \\
\frac{d\sigma_T}{d\alpha dp_\perp^2} & = &\frac{e_c^2\alpha_{em}\alpha_s
\ln(1/x)h_{\cal A}}{16\pi^3(a^2+p_\perp^2)^4}\,\left[(\alpha^2+
(1\!-\!\alpha)^2)\,(p_\perp^4+a^4)+2p_\perp^2m_c^2\right]  \\
h_{\cal A} & = & \int_{y_\perp}\int_{x_\perp}\frac{\left|\mbox{tr}
W^{\cal A}_{x_\perp}(y_\perp)\right|^2}{y_\perp^4} 
\eea
\noindent where $y_{\perp}$ is the transverse separation between  the
gluon and the $c {\bar c}$ pair.

If one boosts this system to the Breit frame, 
the initial state, off-shell, `slow' gluon turns around
and may be reinterpreted as an incoming gluon of momentum fraction $y$
which lies between $\xi$ and $x$ \cite{MFM:AH}. 
In this frame we have boson-gluon fusion off the diffractive gluon density in
the proton \cite{MFM:BS} with an additional final state gluon
(cf. Fig.(\ref{F:MFM:GAP})). The $p^2_\perp$ spectra is then 
logarithmically distributed between $m_c^2$ and $Q^2$. This should be
contrasted to the much softer spectra expected from the two gluon
exchange graphs where typically $p^2_{\perp} \sim m_c^2 $. Note the
above cross sections are also enhanced by a logarithm at small $x$ over
configurations with a wee quark. In addition the constant, 
$h_{\cal A}$, is expected to be considerably bigger than its
equivalent in the fundamental representation 
($h_{\cal A} \approx 16 h_{\cal F}$)  providing a further enhancement
\cite{MFM:BHMC}.

\gnufig{F:MFM:GAP}{Boson-gluon fusion, with a final state gluon, in
  the Breit frame}{
\vspace{-0.7cm}
\hspace{0.8cm}
\epsfig{file=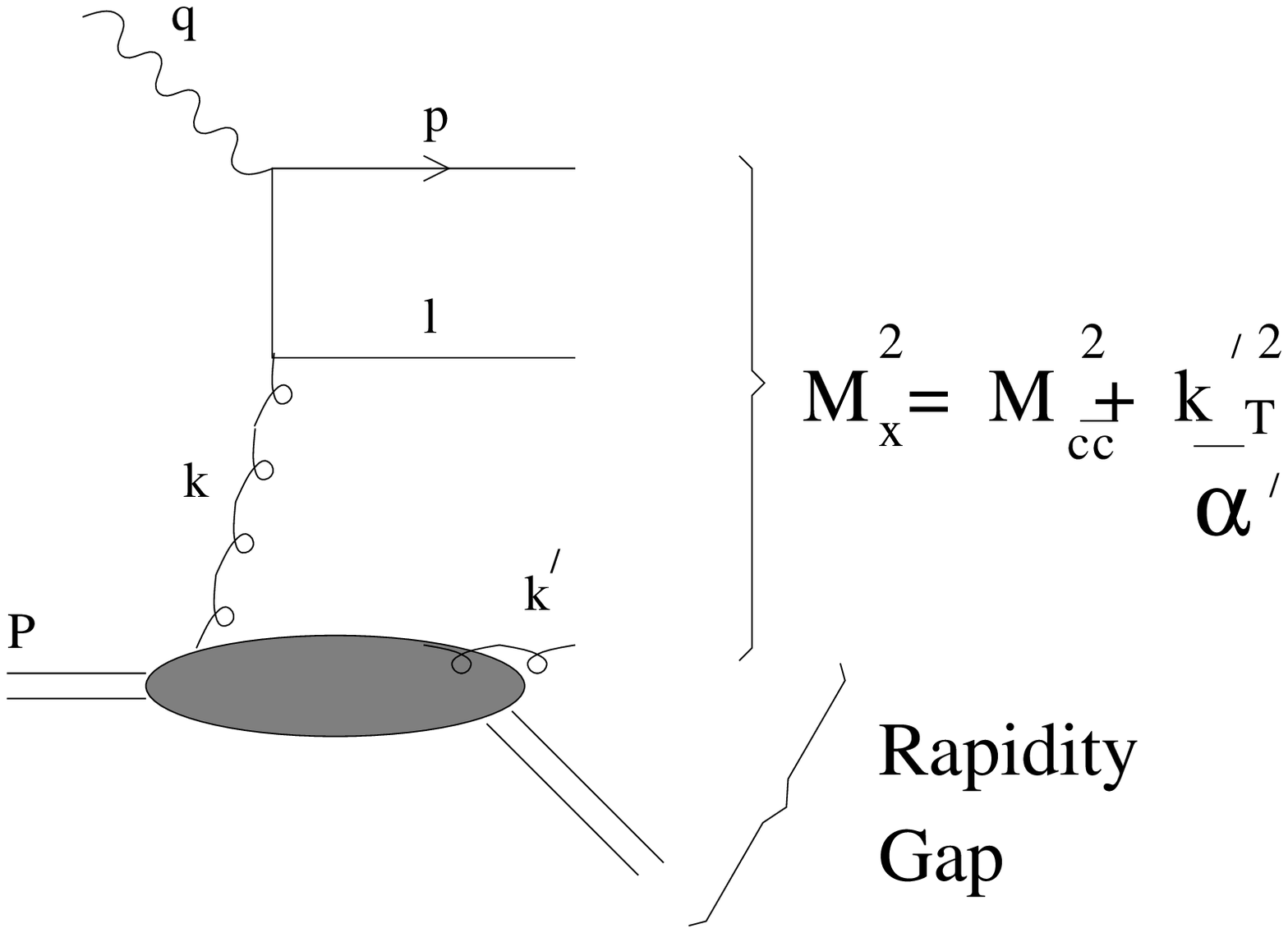,width=8cm}
}

In order to distinguish which type of graph is dominant we propose two
phenomenological tests which reveal the nature of the underlying
process and reflect the mechanism of colour neutralization which
produces the different diffractive final states. Firstly, one may ask
how many diffractive charm events survive above a given minimum
$p_{\perp}^2$. To examine this we plot the quantity 
 \bea
\sigma(p^2_{\perp,\mbox{\footnotesize min}}) &=& 
\int_{p^2_{\perp,\mbox{\scriptsize min}}}^{\infty}
{dp^2_\perp} \int_{0}^{1} d\alpha \frac{d^2 \sigma}{dp_\perp^2
  d\alpha} 
\eea
\noindent in Fig.(\ref{F:MFM:PSQ}). 

\gnufig{F:MFM:PSQ}{
The fraction of diffractive charm events above 
$p^2_{\perp,\mbox{\footnotesize min}}$ for $Q^2$ of 10 GeV$^2$ and 100 
GeV$^2$ (lower and upper curve in each pair).}{
\vspace*{-.7cm}
\setlength{\unitlength}{0.1bp}
\begin{picture}(3600,2160)(0,0)
\includegraphics{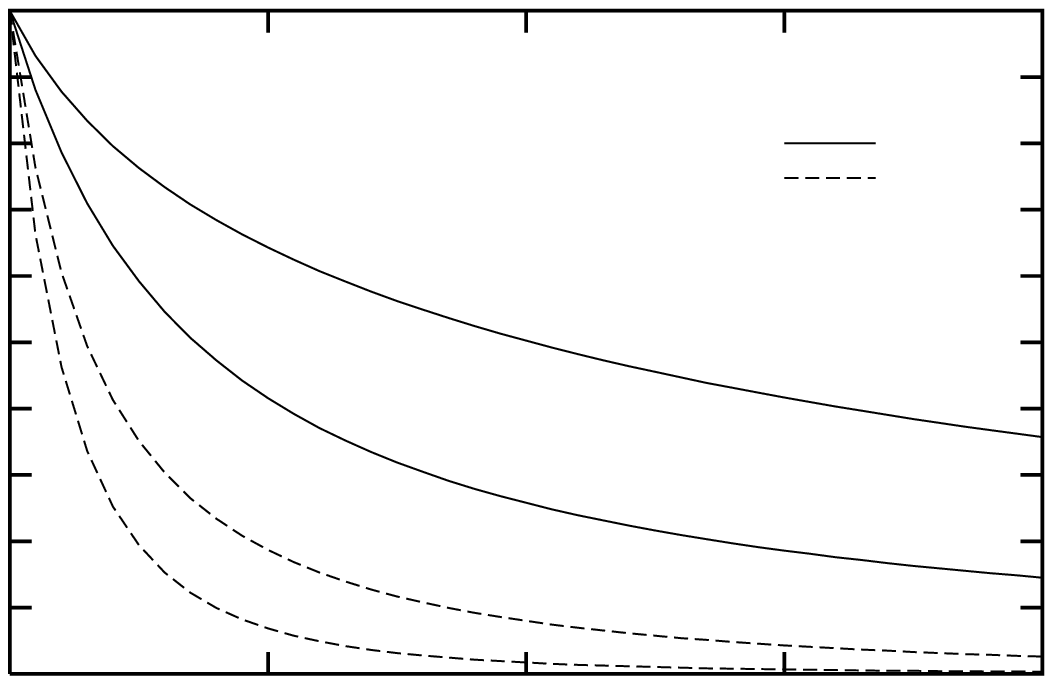}
\put(2619,1578){\makebox(0,0)[r]{two-gluon model ($c {\bar c}$)}}
\put(2619,1678){\makebox(0,0)[r]{semiclassical approach  ($c {\bar c} g$)}}
\put(1650,-70){\makebox(0,0)[l]{$p^2_{\perp,\mbox{\footnotesize min}}
~(\mbox{GeV}^2)$}}
\put(100,1105){%
\makebox(0,0)[b]{\shortstack{Fraction  of Events}}%
}
\put(3437,50){\makebox(0,0){20}}
\put(2694,50){\makebox(0,0){15}}
\put(1950,50){\makebox(0,0){10}}
\put(1207,50){\makebox(0,0){5}}
\put(463,50){\makebox(0,0){0}}
\put(413,2060){\makebox(0,0)[r]{1}}
\put(413,1869){\makebox(0,0)[r]{0.9}}
\put(413,1678){\makebox(0,0)[r]{0.8}}
\put(413,1487){\makebox(0,0)[r]{0.7}}
\put(413,1296){\makebox(0,0)[r]{0.6}}
\put(413,1105){\makebox(0,0)[r]{0.5}}
\put(413,914){\makebox(0,0)[r]{0.4}}
\put(413,723){\makebox(0,0)[r]{0.3}}
\put(413,532){\makebox(0,0)[r]{0.2}}
\put(413,341){\makebox(0,0)[r]{0.1}}
\put(413,150){\makebox(0,0)[r]{0}}
\vspace{1.0cm}
\end{picture}
}

\gnufig{F:MFM:MSQ}{Normalized mass spectra for the $c {\bar c} g$
final state calculated in the semiclassical approach and the $c {\bar c}$
final state calculated in the two gluon exchange models.}{
\vspace*{-1.7cm}
\setlength{\unitlength}{0.1bp}
\begin{picture}(3600,2160)(0,0)
\includegraphics{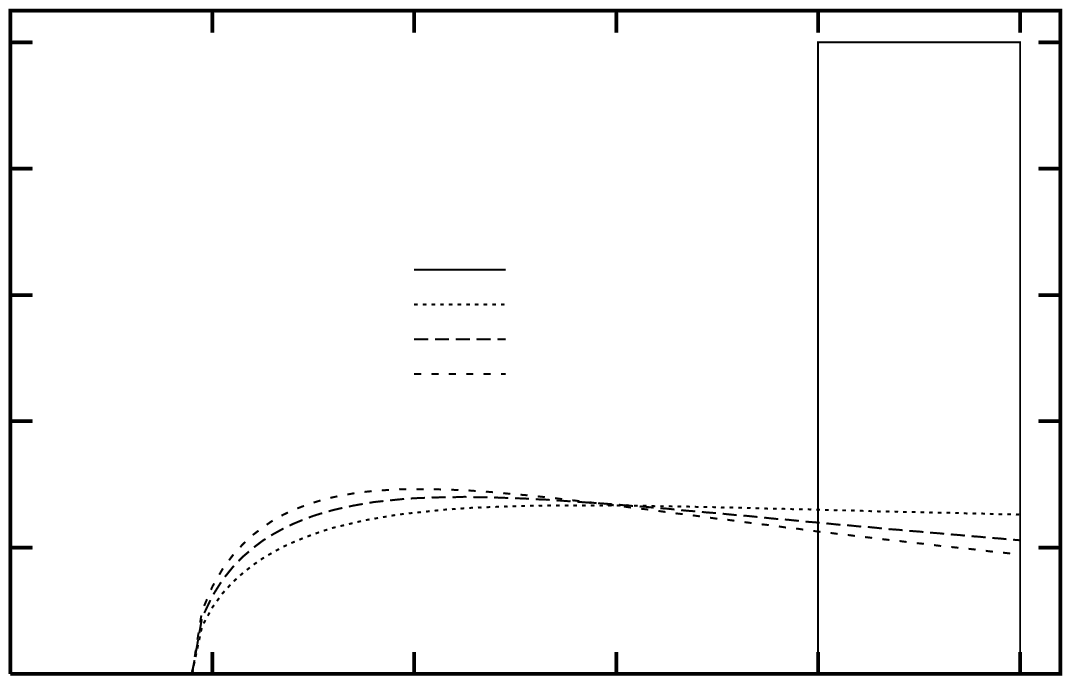}
\put(1526,1014){\makebox(0,0)[r]{$C_s = 2.0$}}
\put(1526,1114){\makebox(0,0)[r]{$C_s = 1.0$}}
\put(1526,1214){\makebox(0,0)[r]{$C_s = 0.5$}}
\put(1526,1314){\makebox(0,0)[r]{$c {\bar c}$ final state}}
\put(1029,1551){\makebox(0,0)[l]{$Q^2 ~= 50 \mbox{~GeV}^2$}}
\put(1751,-70){\makebox(0,0)[l]{$M^2_{c{\bar c}} ~(\mbox{GeV}^2)$}}
\put(-400,1096){\makebox(0,0)[l]{{\Large $\frac{d^2 \sigma_{{\tiny T}}}
{dM^2 dM^2_{c{\bar c}} } $}}}
\put(1038,1442){\makebox(0,0)[l]{$M^2 = 50 \mbox{~GeV}^2$}}
\put(3321,50){\makebox(0,0){50}}
\put(2739,50){\makebox(0,0){40}}
\put(2158,50){\makebox(0,0){30}}
\put(1576,50){\makebox(0,0){20}}
\put(995,50){\makebox(0,0){10}}
\put(413,50){\makebox(0,0){0}}
\put(363,1969){\makebox(0,0)[r]{0.1}}
\put(363,1605){\makebox(0,0)[r]{0.08}}
\put(363,1241){\makebox(0,0)[r]{0.06}}
\put(363,878){\makebox(0,0)[r]{0.04}}
\put(363,514){\makebox(0,0)[r]{0.02}}
\put(363,150){\makebox(0,0)[r]{0}}
\vspace{1.0cm}
\end{picture}
}

We may also examine the expected mass spectra for the different final states.
In Fig.(\ref{F:MFM:MSQ}) a normalized mass spectra comparing the mass of
the charm pair, $M_{c{\bar c}}$, with the total diffractive mass, $M$,
is shown. 
For the $c {\bar c}$ final state of Fig.(\ref{F:MFM:TWOG}) $M_{c{\bar c}} = M$,
up to hadronization effects, and is represented by a block, where the width
of the block represents a guess at the experimental uncertainty of
diffractive mass measurement. The other curves in the figure represent
the $c{\bar c}g$ final state of Figs.(\ref{F:MFM:QQG},\ref{F:MFM:GAP})
for which the mass of the pair may be considerably less than the
total diffractive mass, as a result of the `slow' gluon in the final state.
Details abut how the latter curves were arrived at and the meaning 
of $C_s$ are given in \cite{MFM:BHMC}.

An additional means of distinguishing the underlying diffractive
mechanism is the energy dependence of the two processes. In the two
gluon model a steeply rising gluon density, taken from a fit to $F_2$
for example, produces a rise in
diffractive charm events that is twice as steep with energy. 
In contrast the semiclassical approach the energy dependence 
is flat, at least at this order,
corresponding to a classical bremstrahl spectrum of gluons.

Given the increase in statistics of Hera for the 1996 running period, it
is hoped that these phenomenological tests may be performed very soon.


\begin{references}

\bibitem{MFM:H1WAR}
H1 collab., pa02-060, {\it A Measurement of the Production of 
$D^{*\pm}$ Mesons in Deep-Inelastic Diffractive Interactions at HERA},
XXVIII ICHEP, Warsaw, 1996

\bibitem{MFM:DATCHI}
C. Cormack, H1 collab., these proceedings; 
J Terron, ZEUS collab., these proceedings.

\bibitem{MFM:TWOG}
E.M. Levin, A.D. Martin, M.G. Ryskin, and T. Teubner, hep-ph/9606443;
M. Genovese, N.N. Nikolaev, B.G. Zakharov, Phys. Lett. B378 (1996) 347;
H. Lotter, preprint DESY 96-260, hep-ph/9612415;
M. Diehl, preprint CPTH-S492-0197, hep-ph/9701252

\bibitem{MFM:BHM}
W. Buchm\"uller and A. Hebecker, Nucl. Phys. B476 (1996) 203;
W. Buchm\"uller, M.F. McDermott, and A. Hebecker, Nucl. Phys. B487 (1997) 283, erratum {\it ibid.}

\bibitem{MFM:BHMC}
W. Buchm\"uller, M.F. McDermott, and A. Hebecker, hep-ph/9703314.

\bibitem{MFM:AH}
A. Hebecker, preprint DAMTP-97-10, hep-ph/9702373.

\bibitem{MFM:BS}
A. Berera and D. E. Soper , Phys Rev D53 (1996) 203; Phys Rev D50
(1994) 4328.

\end{references}
\end{document}